\documentclass[twocolumn]{webofc}
\usepackage[varg]{txfonts}   
\def\Xmax{X_{\rm max}}
\begin{document}
\title{Air showers, hadronic models, and muon production}
%
%

\author{\firstname{S. J.}
  \lastname{Sciutto}\inst{1}\fnsep\thanks{\email{sciutto@fisica.unlp.edu.ar}}
}

\institute{IFLP CONICET and Departamento de F{\'\i}sica,
           Universidad Nacional de La Plata,
           C. C. 67 -- 1900 La Plata, Argentina
          }

\abstract{%
 We report on a study about some characteristics of muon
  production during the development of extended air showers initiated
  by ultra-high-energy cosmic rays. Using simulations with the recent new
  version of the AIRES air shower simulation system, we analyze and
  discuss on the observed discrepancies between experimental
  measurements and simulated data.
}
\maketitle
\section{Introduction}
\label{sec:intro}

The determination of the composition of the ultra-energetic cosmic
rays is one of the most relevant current challenges of Cosmic Ray
experimental physics. The hypothesis widely accepted at present is
that the flux of ultra-energetic cosmic rays incident on the
terrestrial atmosphere is constituted essentially by a mixture of
different atomic nuclei. Under this hypothesis, the determination of
the primary composition is equivalent to the determination of the
masses of the incident nuclei. In the experiments that can measure the
longitudinal development of particle showers initiated after the
incidence of the mentioned ultra-energetic cosmic rays, the
atmospheric depth of the maximum of these showers, $\Xmax$, is a robust
observable directly correlated with the mass of the primary
particle. Such correlation can be established by comparison with
numerical simulations that prove to depend on the
algorithms used to model the hadronic collisions that take place
during shower development \cite{LpHadModels2018}.

Measurements of $\Xmax$ and its fluctuations made by the Pierre Auger
Observatory \cite{AugerXmaxI2014,AugerXmaxICRC2017} seem to indicate
that the composition of ultra-energetic cosmic rays varies with their
energy. The results published in
\cite{AugerXmaxI2014,AugerXmaxICRC2017} clearly show that the values
of $\Xmax$ and its fluctuations are approximately compatible with
those obtained by numerical simulations for light primaries ($A \sim
1$) for primary energies around $2\times 10^{18}$ eV, while for larger
energies a progressive shift towards a heavier composition is
observed. This characteristic of variable composition with energy is
analyzed in several works. In particular, in \cite{AugerCombFit} an
adjustment of the total flux of cosmic rays as a function of primary
energy is made, simultaneously with $\Xmax$ and its fluctuations,
using a conveniently parameterized procedure, obtaining a combined
setting that describes the total flux of ultra-energetic cosmic rays
that arrive at the Earth as a mixture of elements (H, He, N, and Si)
obtaining for each one of them the corresponding partial contribution
in function of the primary energy. This work also includes a
discussion on extensions of the fitted contibutions for energies below
$5\times 10^{18}$ eV that adds a fifth contribution assigned to Fe
nuclei (see discussion in reference \cite{AugerCombFit}).

\begin{figure*}
\centering
\includegraphics[width=10cm,clip]{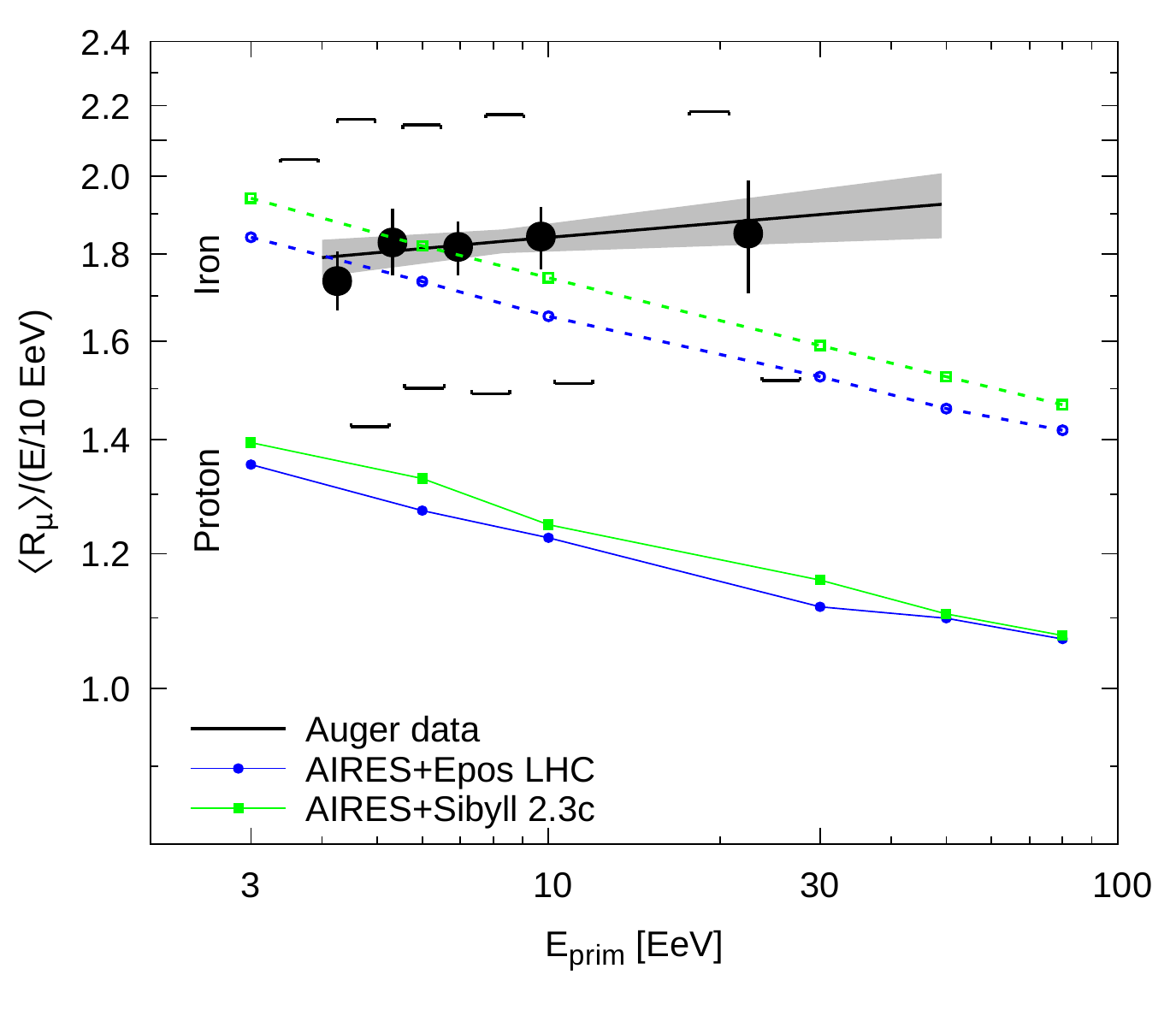}
\caption{$\langle R_\mu\rangle/E_{\rm prim}$ versus primary energy (I). The black
  circles, straight line with grey shaded zone, and square horizontal
  brackets correspond to data published by the Pierre Auger
  Observatory in reference \cite{AugerInclMu2015} (figure 4). Circles
  error bars and grey shaded zone indicate statistical uncertainties,
  while the horizontal brackets stand for systematic ones. The blue
  and green lines correspond to estimations of $R^{\rm MC}_\mu$
  obtained from simulations with AIRES, linked to the hadronic models
  indicated in the graph, following the procedure described in section
  \ref{sec:simres}.
}
\label{fig:RmuVsEprim1}
\end{figure*}
The muonic content of showers initiated by cosmic rays is another
important observable directly related to the composition of cosmic
rays. However, the estimation of the primary composition of particle
showers using experimental measurements has not yet been
satisfactorily completed due to the discrepancies that are observed
when comparing the experimental data with the results of computational
simulations \cite{AugerInclMu2015,AugerHadModMu2016,TAMu2018}. An
example of this is the estimation of the number of muons in inclined
showers published by the Pierre Auger Observatory
\cite{AugerInclMu2015}, where it can be clearly observed that the
simulations produce estimates that are significantly lower than the
experimental results. It is important to mention that similar
disagreements between experimental measurements and simulations are
also obtained with alternative analyses such as those reported in
references \cite{AugerHadModMu2016,TAMu2018}.

In this work we address the issue of the observed
discrepancies between experiment and simulations of the muon
production in air showers by presenting an alternative discussion of
the estimates of the number of muons at ground level, under conditions
similar to those of reference \cite{AugerInclMu2015}, using
simulations carried out with the new version of our AIRES program,
recently publicly launched \cite{AIRES2001,AIRES2018}.

\section{Simulations and results}
\label{sec:simres}

We recall that AIRES \cite{AIRES2001,AIRES2018} is a system for the
realistic simulation of particle showers in the Earth's atmosphere,
also containing a series of tools for the analysis of the
corresponding results. In its current version, 18.09.00, recently
launched \cite{AIRES2018}, a large series of new features have been
incorporated. Among these new features, we find important to mention:
(1) Links with the hadronic interaction models EPOS \cite{EPOS},
QGSJET \cite{QGSJET}, and SIBYLL \cite{SIBYLL}, in both their pre- and
post-LHC versions. (2) Propagation of an extended set of particles,
including, for example, tau leptons, and charmed and bottom
hadrons. (3) Detailed simulation of decays of unstable particles,
including up to many tens of relevant branches in the case of particles
with complex decay patterns. (4) Extended sets of data that are
recorded in the listings of particles reaching the ground or of
longitudinal survey, including, for example, identity and energy of
the mother particle, identity and energy of the projectile of the last
hadronic event in the history of the particle, etc. A complete
description of the AIRES system, including access to the corresponding
user manual, is placed at \cite{AIRES2018}.

As a first step of our analysis with AIRES, we have carried out
numerous simulations in order to reproduce the most relevant results
that have already been published. In the case of $\Xmax$ and its
fluctuations, we have verified that there is an excellent agreement
between the simulations with the current version of AIRES and those
published in references \cite{AugerXmaxI2014,AugerXmaxICRC2017}, which
were made with CONEX \cite{CONEX}, with links to the already mentioned
EPOS, QGSJET, and SIBYLL hadronic models in their pre and post-LHC
versions. For reasons of brevity, we omit presenting in this work the
data in graphic form.

A similar check can be made in the case of the number of muons at
ground level. Following reference \cite{AugerInclMu2015}, we run
simulations of air showers with an inclination from the vertical of 67
degrees, and for each set of showers we evaluate the average number of
muons at ground level that have a total energy larger than 300 MeV,
$N^{\rm MC}_\mu$. Then we calculate $R^{\rm MC}_\mu = N^{\rm
  MC}_\mu/N^{\rm Ref}_\mu$, where $N^{\rm Ref}_\mu = 1.455 \times
10^7$. $R^{\rm MC}_\mu$ can be directly compared with the experimental
measurements, as explained in reference \cite{AugerInclMu2015}.

In figure \ref{fig:RmuVsEprim1} we present our results for
 $R^{\rm MC}_\mu/E_{\rm prim}$ corresponding to a first series of simulations
performed with AIRES linked to EPOS-LHC or SIBYLL 2.3c, and proton or
iron nuclei primaries. Experimental data of the Pierre Auger
Observatory published in \cite{AugerInclMu2015} is also displayed for
comparison purposes. We verified that the simulations performed with
AIRES + EPOS-LHC (blue lines) are virtually coincident with those
corresponding to simulations with CORSIKA \cite{CORSIKA} + EPOS-LHC
plotted in the similar figure 4 of reference \cite{AugerInclMu2015}.
The green lines in figure \ref{fig:RmuVsEprim1},
correspond to simulations with AIRES + SIBYLL 2.3c, and correspond to larger
values of $R^{\rm MC}_\mu$ than the CORSIKA + SIBYLL 2.1 values
represented in figure 4 of the aforementioned
reference \cite{AugerInclMu2015}, as expected for this pre-LHC version of
SIBYLL which produces a smaller number of muons during the development
of the showers (for a detailed discussion on this issue see reference
\cite{LpHadModels2018}).

\begin{figure}
\centering
\includegraphics[clip]{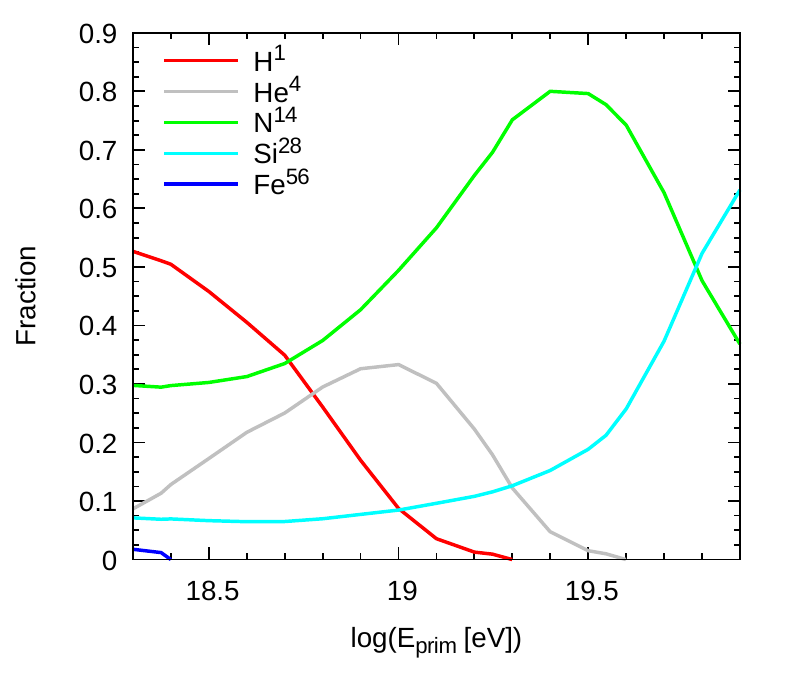}
\caption{Fractions of ultra-high energy primary cosmic rays entering
  at the top of the Earth's atmosphere, as functions of the primary
  energy, evaluated from partial fluxes corresponding to the combined
  fit published by the Pierre 
  Auger Observatory \cite{AugerCombFit} (see text).
}
\label{fig:CmbFitFracs}       
\end{figure}
In both simulations shown in figure \ref{fig:RmuVsEprim1} it is
evident that the number of muons predicted by them is significantly
lower than the corresponding experimental data. As already mentioned,
this lack of agreement between experimental data and simulations is
presently one of the most relevant enigmas of ultra-energetic cosmic
rays, and is receiving most attention in current research. It
should be noted that this "excess" of muons in the experimental data
is also reported in other analysis, either from the Pierre Auger
Observatory \cite{AugerHadModMu2016}, or the Telescope Array
Collaboration \cite{TAMu2018}.

\begin{figure*}[p]
\centering
\begin{tabular}{cc}
\includegraphics[width=8cm,clip]{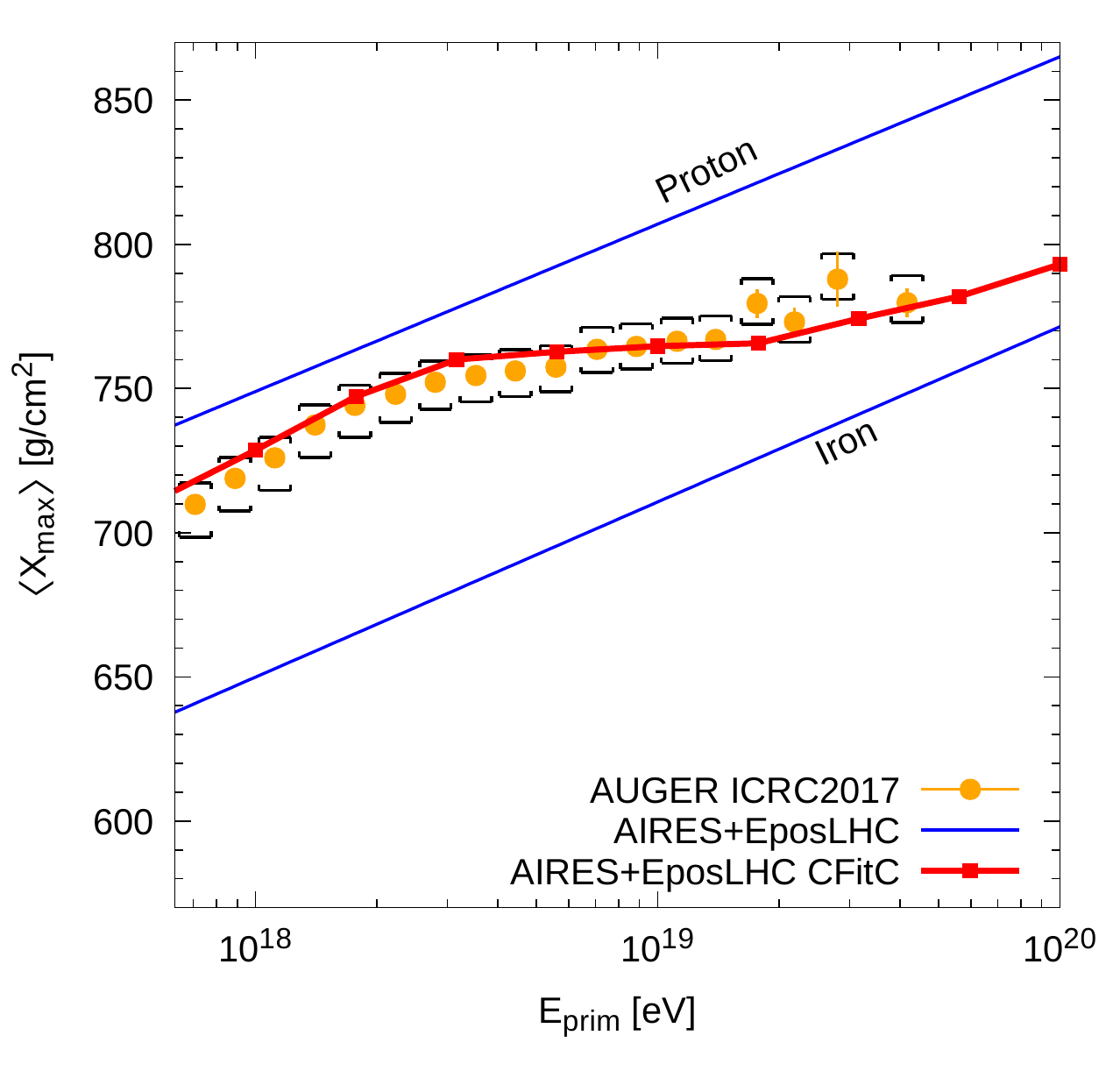}
&
\includegraphics[width=8cm,clip]{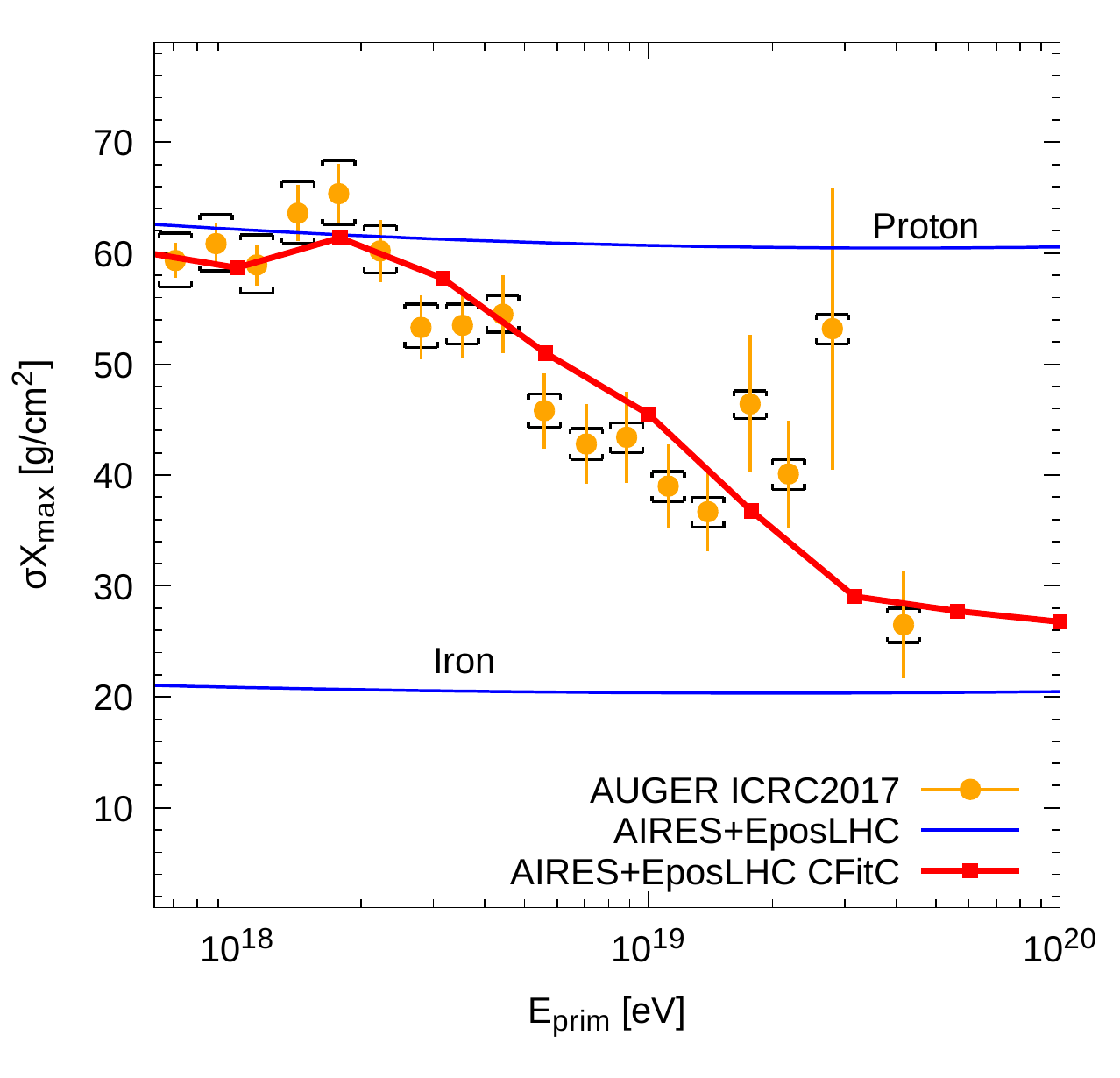}
\end{tabular}
\caption{$\Xmax$ and $\sigma\Xmax$ versus primary energy. The experimental
  points reproduce the Auger Observatory data published in reference
  \cite{AugerXmaxICRC2017}. The error bars (horizontal square
  brackets) correspond to statistical (systematic) uncertainties. The
  red curves (labelled CFitC) correspond to simulations with AIRES +
  EPOS-LHC using an energy dependent combined composition (see text).}
\label{fig:XmaxSXmaxVsEprim}
\end{figure*}
\begin{figure*}[p]
\centering
\includegraphics[width=10cm,clip]{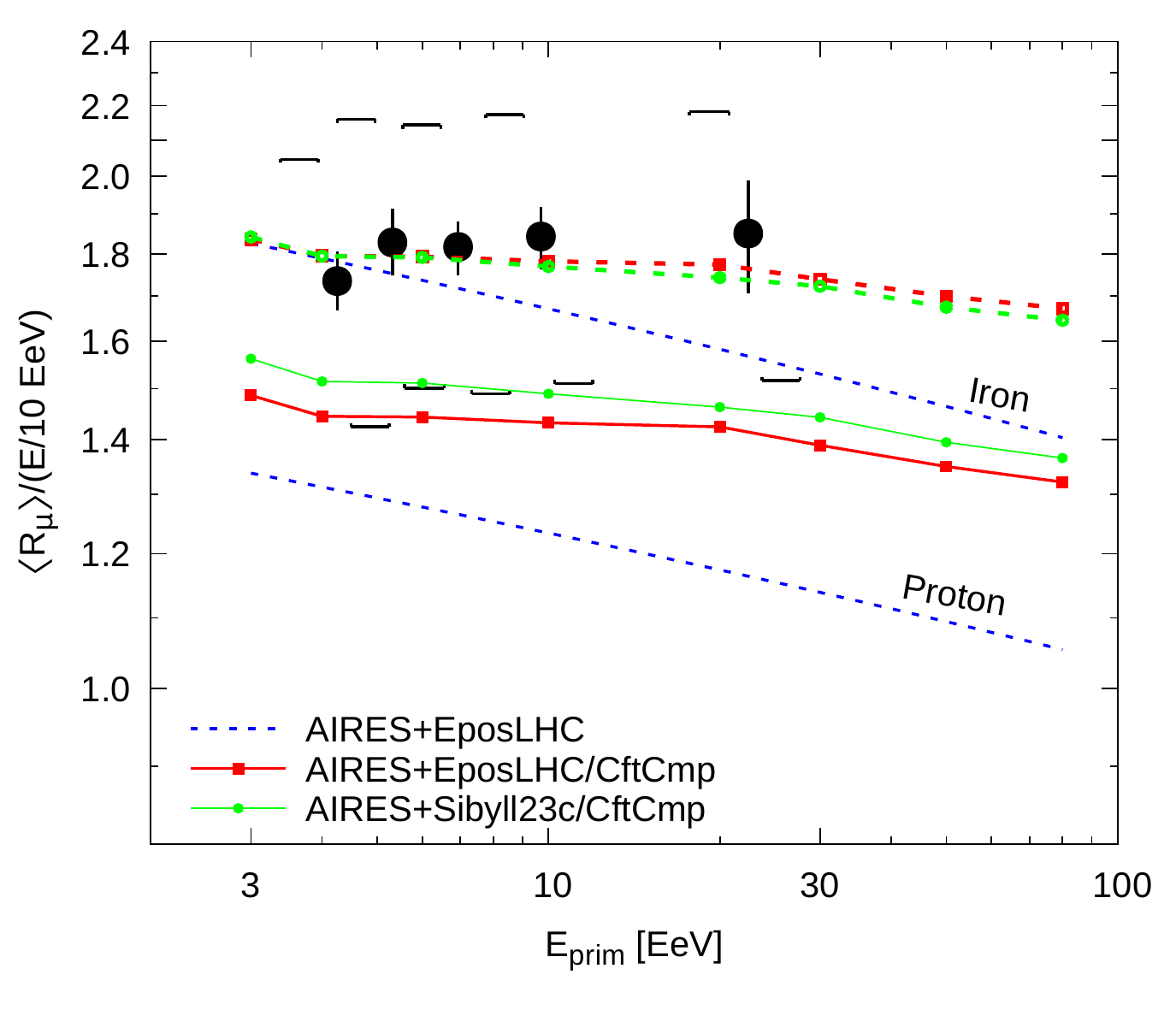}
\caption{$\langle R_\mu\rangle/E_{\rm prim}$ versus primary energy (II). The
  black circles and square horizontal brackets correspond to data
  published by the Pierre Auger Observatory in reference
  \cite{AugerInclMu2015} (figure 4). Circles error bars indicate
  statistical uncertainties, while the horizontal brackets stand for
  systematic ones. The dashed blue lines correspond to least square
  fits to estimations of $R^{\rm MC}_\mu$ obtained from simulations
  with fixed composition. The red squares (green circles) with solid
  lines (labelled CftCmp) correspond to estimations of $R^{\rm
    MC}_\mu$ obtained with AIRES, linked to the hadronic models
  indicated in the graph, and using an energy dependent combined
  composition (see text). The dashed curves represent the
  corresponding solid line ones, shifted by a given constant offset
  adjusted to match the experimental data points. All the plotted
  values of $R^{\rm MC}_\mu$ were calculated using simulations with
  AIRES and following the procedure described in section
  \ref{sec:simres}.
}
\label{fig:RmuVsEprimCmbFit}       
\end{figure*}

The curves of figure \ref{fig:RmuVsEprim1} corresponding to the
simulations for proton and iron clearly show the well-known
characteristic that for showers of a given primary energy, the
production of muons is larger the larger the mass of the
primary. Connecting this with the already mentioned result that
measurements of the depth of the shower maximum, $\Xmax$, and
its fluctuations provide favorable evidence for a mixed primary
composition of cosmic rays that reach the Earth's atmosphere,
with proportions that depend on energy, we consider it worthwhile to
investigate what the impact of this hypothesis would be
in the case of muon production.

To this end, we take into account the adjustment published in
reference \cite{AugerCombFit}, where the flux of cosmic rays with
primary energies greater than $5\times 10^{18}$ eV is the sum of four
components, namely, protons (H$^1$), He$^4$, N$^{14}$, and Si$^{28}$,
whose relative abundance is estimated with an elaborate procedure,
described in detail in reference \cite{AugerCombFit}, which takes into
account the whole process of creation, acceleration and propagation of
the primaries through the intergalactic space, and allows the
simultaneous adjustment of the energy spectrum, $\Xmax$, and its
fluctuations $\sigma\Xmax$. This combined adjustment can be extended
to energies slightly below $5\times 10^{18}$ eV by adding an
additional contribution to the cosmic ray flux, of galactic origin and
including a new heavy fraction that is considered to be composed of
iron (Fe$^{56}$) nuclei (see reference \cite{AugerCombFit}, figure 14
and the related discussion). From these results published by the
Pierre Auger Observatory, we calculated the normalized abundances of
the five components, protons (H), He, N, Si and Fe, in the range of
energies relevant to our study of muon production in inclined showers,
which are represented in figure \ref{fig:CmbFitFracs}.

With these primary fractions plotted in figure \ref{fig:CmbFitFracs}
we have carried out numerous simulations with AIRES. First, we have
verified the correct adjustment of $\Xmax$ and its fluctuations with
the results published by the Pierre Auger Observatory
\cite{AugerXmaxI2014,AugerXmaxICRC2017}, which are also used for the
combined adjustment of the reference \cite{AugerCombFit}. Our results
are shown in figure \ref{fig:XmaxSXmaxVsEprim} (curves in red line),
where the very good degree of agreement between the simulations and
the experimental values of both $\Xmax$ and its fluctuations can be
appreciated.

Finally we have carried out simulations to determine $R^{\rm MC}_\mu$
under the hypothesis of a combined composition, with the abundances as
represented in figure \ref{fig:CmbFitFracs}. The results are shown in
figure \ref{fig:RmuVsEprimCmbFit}. The red squares with solid lines
represent the results with AIRES + EPOS-LHC, this being the same
hadronic model as used in the combined fit of reference
\cite{AugerCombFit}. The most outstanding aspect of this curve is that
it is approximately horizontal, that is, with a similar dependence on
energy as the experimental data points. Thus, the difference between
simulations and experimental data virtually reduces to a constant that
accounts for the muon deficit that arises from comparing the
experimental data with the simulations. In effect, if we add a
constant, properly adjusted, to the values of $R^{\rm MC}_\mu/
 E_{\rm prim}$ of this red curve, we can raise it and make it coincide with
all the experimental points within the statistical errors range
(red dashed curve figure \ref{fig:RmuVsEprimCmbFit}).

Figure \ref{fig:RmuVsEprimCmbFit} also includes the results of the
simulations with AIRES + SIBYLL 2.3c (green circles with solid line)
which show a behavior similar to the case of EPOS-LHC, with slightly
higher muon production. As in the case of simulations with EPOS-LHC,
the curve corresponding to SIBYLL 2.3c can be moved towards higher
values of $R^{\rm MC}_\mu/ E_{\rm prim}$ with the addition of a
global constant, and get a one sigma adjustment to all experimental
data points.

\section{Final remarks}
\label{sec:final}

A new study of the production of muons in showers initiated by
ultra-energetic cosmic rays has been presented, using simulations with
the recently released version of AIRES \cite{AIRES2018}.

To carry out this study it has been necessary to verify the proper
simulation of usual observables, such as $\Xmax$ and its fluctuations, for
example, in a series of well-known cases, allowing to ensure the
correctness of the new interfaces with hadronic models that
are currently included in AIRES.

In the case of the number of muons in inclined showers, it has been
possible to corroborate that the simulations with AIRES linked to
the EPOS-LHC, QGSJET-II-04, and SIBYLL 2.3c models, predict a
significantly lower number of muons compared to the experimental data,
in the same direction as the results reported in the references
\cite{AugerInclMu2015,AugerHadModMu2016,TAMu2018}.

In our analysis of the number of muons, we have also made the
comparison of the experimental data with simulations taking into
account a combined primary composition \cite{AugerCombFit}, with
abundances dependent on energy. In this case, it is noteworthy that
although a muon deficit persists in the simulated data with respect to
the experimental ones, this difference can be virtually reduced to a
constant, independent of the primary energy, shift in $R_\mu/E$ which
added to the results of the simulations allows to locate the displaced
curve in agreement with the experimental data within error bars. We
understand this as a new indication supporting the hypothesis of the
combined composition of ultra-energetic cosmic rays. The results of
this analysis call for a more complete study in order to understand
the origin of this displacement, and to improve the hadronic models to
obtain predictions compatible with the experimental results. We are
currently investigating in this direction.

\section*{Acknowledgments}

The author is indebted to C. A. Garc{\'\i}a Canal, L. Anchordoqui,
L. Calcagni, and T. Tarutina, for enlightening discussions on the
mechanisms for muon production in air showers.
This work was partially supported by ANPCyT and CONICET, Argentina.

%

\begin{thebibliography}{99}
%
%
\bibitem{LpHadModels2018}
L. Calcagni, C. A. Garc{\'\i}a Canal, S. J. Sciutto, T. Tarutina,
Phys. Rev. D \textbf{98}, 083003 (2018).
%
\bibitem{AugerXmaxI2014}  A. Aab \textit{et al.} (The Pierre Auger
  Collaboration), Phys. Rev. D \textbf{90}, 122005 (2014).
%
\bibitem{AugerXmaxICRC2017} J. Bellido for The Pierre Auger
  Collaboration, Proc. of the 35th Int. Cosmic Ray
  Conf., ICRC 2017 (Bexco, Busan, Korea), PoS(ICRC2017) 506.
%
\bibitem{AugerCombFit}  A. Aab \textit{et al.} (The Pierre Auger
  Collaboration), JCAP \textbf{04} 038 (2017); erratum: JCAP
  \textbf{03} E02 (2018).
%
\bibitem{AugerInclMu2015} A. Aab \textit{et al.} (The Pierre Auger
  Collaboration), Phys. Rev. D \textbf{91}, 032003 (2015).
%
\bibitem{AugerHadModMu2016} A. Aab \textit{et al.} (The Pierre Auger
  Collaboration), Phys. Rev. Lett. \textbf{117}, 192001 (2016).
%
\bibitem{TAMu2018} R. U. Abbasi \textit{et al.} (The Telescope Array
  Collaboration), Phys. Rev. D \textbf{98}, 022002 (2018).
%
\bibitem{AIRES2001} S. J. Sciutto, Proc. 27th ICRC (Hamburg)
  \textbf{1}, 237 (2001).
%
\bibitem{AIRES2018} Updated information about AIRES can be found at
  the site {\em aires.fisica.unlp.edu.ar}
%
\bibitem{EPOS}
T. Pierog, Iu. Karpenko, J.M. Katzy, E. Yatsenko, K. Werner,
Phys. Rev. C \textbf{92}, 034906 (2015).
%
\bibitem{QGSJET}
S. S. Ostapchenko, Phys. Rev. D \textbf{83}, 014018 (2011).
%
\bibitem{SIBYLL}
  Eun-Joo Ahn \textit{et al.}, Phys. Rev. D \textbf{80}, 094003 (2009);
  F. Riehn \textit{et al.}, Proc. 35th Int. Cosmic Ray Conf., ICRC 2017
  (Bexco, Busan, Korea), PoS(ICRC2017) 301.
%
\bibitem{CONEX}
T. Bergmann, R. Engel, D. Heck, N. N. Kalmykov, S. Ostapchenko, T. Pierog,
T. Thouw, K. Werner,
Astroparticle Physics, \textbf{26}, 420 (2007).
%
\bibitem{CORSIKA}
D. Heck, G. Schatz, T. Thouw, J. Knapp, and J. N. Capdevielle, Report
No. FZKA 6019 (1998).
%
\end{thebibliography}
%
%

\end{document}